\documentclass[10pt]{amsart}

\usepackage{stmaryrd}
\usepackage{amsfonts}
\usepackage{bbm}
\usepackage{mathrsfs}
\usepackage{latexsym,amssymb,amsmath,amscd,amsthm,amsxtra}
\usepackage[utf8]{inputenc}
\usepackage[T1]{fontenc}
\usepackage{nicefrac,mathtools,enumitem}
\usepackage{microtype}
\usepackage{changepage}

\usepackage{tikzsymbols}

\newenvironment{pf}{\paragraph{{\sc Proof}}}{\qed\par\medskip}
\newtheorem{thm}{Theorem}[section]

\newtheorem{pro}[thm]{Proposition}

\newtheorem{defi}[thm]{Definition}

\newcommand{\be }{\begin{equation}}
\newcommand{\ee }{\end{equation}}

\newcommand{\h}{\mathbbm h}

\newcommand{\br}[1]{   [ \cdot,    \cdot  ]   }

\newcommand{\g}{\mathfrak g}

\newcommand {\IR}{\mathbb{R}}

\renewcommand {\h}{\mathfrak h}

\newcommand {\comment}[1]{\footnote{\textcolor{blue}{#1}}}
\renewcommand {\comment}[1]{}
\newcommand {\Omit}[1]{}

\newcommand {\IH}{\mathbb H}

\title{Stokes phenomenon and Yang-Baxter equations}
\author{\small XIAOMENG XU}
\date{}
\newcommand{\Addresses}{{
  \bigskip
  \footnotesize

  \textsc{DEPARTMENT OF MATHEMATICS, MASSACHUSETTS INSTITUTE OF TECHNOLOGY, CAMBRIDGE, MA 02139, USA}\par\nopagebreak
  \textit{E-mail address}: \texttt{xxu@mit.edu}

}}

\begin{document}
\maketitle
\begin{abstract}
We describe the monodromy of dynamical Knizhnik-Zamolodchikov equations via Stokes phenomenon. It defines a family of braid groups representations by certain Stokes matrices. In particular, these Stokes matrices satisfy the Yang-Baxter equation.
\end{abstract}
\section{Introduction}
The Knizhnik-Zamolodchikov (KZ) equation \cite{KZ} is a local system on the configuration space of points. Its monodromy produces a braid group representation, and is closely related to conformal field theory, quantum groups, representation theory of affine Lie algebras, geometry of cycles and so on. See e.g., Drinfeld \cite{Drinfeld}, Kohno \cite{Kohno}, Kazhdan-Lusztig \cite{KL}, Varchenko \cite{Va}, Etingof-Varchenko \cite{EV} and the reference therein.

The dynamical KZ ($\rm dKZ$) equation, introduced by Felder, Markov, Tarasov and Varchenko \cite{FMTV} in the study of bispectral problem, is a $\rm KZ$ type equation coupled with irregular singularities. The purpose of this paper is to study its monodromy using resummation methods (known as the basic tools for studying equations with singularities, see e.g., Balser \cite{Balser} and Malgrange-Ramis \cite{MR}). Along the way, we show that the Yang-Baxter equation, arising in quantum field theory and statistical mechanics, can be understood within the framework of the Stokes phenomenon (the discontinuity of the asymptotics as one approaches to the irregular singularity from different directions). The Stokes phenomenon of ${\rm dKZ}$ equations in the formal setting was used by Toledano Laredo \cite{TL} to construct Drinfeld twists killing KZ associators, and was shown to be related to R-matrices, Poisson Lie groups, Gelfand-Zeitlin systems, and the theory of Frobenius manifolds in our works \cite{TLXu} \cite{Xu}.

\subsection{Monodromy representation of the ${\rm dKZ}$ equations}\label{intro}
In this paper, we shall be concerned with the ${\rm dKZ}_n$ equation associated with ${\rm gl}_m$, the natural representation $V\cong\mathbb{C}^m$ and a diagonal matrix $u\in{\rm gl}_m$ with distinct diagonal elements. It is an equation for a $V^{\otimes n}$-valued function $F(z_1,...,z_n)$ of $n$ complex variables
\begin{eqnarray}
\kappa\frac{{\partial F }}{\partial z_i}= (u^{(i)}+ 
\sum_{j\ne i}\frac{ \Omega_{ij}}{z_i-z_j})\cdot F, \ \ \ \ \ \ i=1,...,n.
\end{eqnarray}
Here $\Omega_{ij}$ and $u^{(i)}$ denote $\Omega$ acting on the $i$-th and $j$-th factors and $u$ acting on the $i$-th factor of $V^{\otimes n}$, and $\Omega=\sum_{1\le a,b\le m} E_{ab}\otimes E_{ba}\in {\rm End}(V)^{\otimes 2}$ for $E_{ab}$ being the elementary matrix whose $ij$-entry is $\delta_{ia}\delta_{jb}$. Our main result can be stated as follows.

\begin{thm}\label{mainthm}
The monodromy representation of the ${\rm dKZ}_n$ equation is given by
$$B_n\rightarrow {\rm End}(V^{\otimes n}); \ b_i\mapsto T_i\circ R^{i,i+1}, \ \ \ \ \ i=1,...,n-1, $$
where $\{b_i\}'s$ are the generators of braid group $B_n$ (see Section \ref{secbraid}), $R^{i,i+1}$ is the action of the Stokes matrix $R\in {\rm End}(V)^{\otimes 2}$ of the ${\rm dKZ}_2$ equation (formula \eqref{Stokesfactor}) on the $i$-th and the $(i+1)$-th factors of $V^{\otimes n}$, and $T_i: V^{\otimes n}\rightarrow V^{\otimes n}$ is the permutation of the $i$-th and the $(i +1)$-th factors.
\end{thm}

This theorem unveils a strong factorization property of the monodromy of ${\rm dKZ}$ equation, see more discussion in Section \ref{advantages}. In the following, we introduce the Stokes matrix $R$ of the ${\rm dKZ}_2$ equation appeared in the theorem.

\subsection{Stokes matrices and Yang-Baxter equations} First the ${\rm dKZ}_2$ equation with two variables is equivalent to an ordinary differential equation. This is because any solution of ${\rm dKZ}_2$ can be written as $F(z_1,z_2)=e^{\frac{z_2}{\kappa}(u^{(1)}+u^{(2)})}Y(z_1-z_2)$, where the $V^{\otimes 2}$-valued function $Y(z)$ satisfies
\begin{eqnarray}\label{eq}
\kappa\frac{dY}{dz}=(u^{(1)}+\frac{\Omega_{12}}{z})\cdot Y.
\end{eqnarray}
Here recall $\Omega_{12}=\sum E_{ab}\otimes E_{ba}$, $u^{(1)}=u\otimes 1\in {\rm End}(V)^{\otimes 2}$, and $u={\rm diag}(u_1,...,u_n)$ with distinct diagonal elements. We assume henceforth that the diagonal elements of $\frac{1}{\kappa}u$ are purely imaginary. 

The equation \eqref{eq} has an irregular singularity at $z =\infty$ of Poincar$\rm\acute{e}$ rank $1$.
It is known from the theory of meromorphic linear systems (see e.g., \cite{Balser}\cite{BJL} or the Appendix) that it has a unique formal power series fundamental solution $\hat{Y}(z)\in {\rm End}(V)^{\otimes 2}$ around $z=\infty$, which will resum to a canonical holomorphic solution $Y_+(z)$ (resp. $Y_-(z)$) in the right half plane $\IH_+$ (resp. in the left half plane $\IH_-$). The discontinuity of the two solutions $Y_\pm$ is measured by the {\it Stokes matrices} $S_\pm\in{\rm End}(V)^{\otimes 2}$, which are determined by
\[Y_-=Y_+\cdot S_+ , \ \ \ \ \  
Y_+=Y_-\cdot e^{\frac{2\pi i}{\kappa} [\Omega]} \cdot S_-
\]
where $[\Omega]:=\sum E_{aa}\otimes E_{aa}$, and the first (resp. second) identity is understood to hold in $\IH_-$
(resp. $\IH_+$) after $ Y_+$ (resp. $ Y_{-}$)
has been analytically continued counterclockwise. Here the factor $e^{\frac{2\pi i}{\kappa} [\Omega]}$ is known as the formal monodromy. 

Taking into account the half formal monodromy, we define the Stokes multiplier
\begin{eqnarray}\label{Stokesfactor}
R=e^{\frac{\pi i}{\kappa} [\Omega]}S_+\in {\rm End}(V)^{\otimes 2}.
\end{eqnarray}
By the equivalence, this is actually the Stokes matrix of the ${\rm dKZ}_2$ equation. As a corollary of Theorem \ref{mainthm}, the Stokes matrix of ${\rm dKZ}_2$ will produce a solution of Yang-Baxter equation. That is

\begin{thm}\label{StokesYB}
The Stokes multiplier $R$ satisfies the Yang-Baxter equation, i.e.,\footnote{
The convention is that if we write $R=\sum X_a\otimes Y_a$, then $R^{12}:=\sum X_a\otimes Y_a\otimes 1, \ \ R^{13}:=\sum X_a\otimes 1\otimes Y_a, \ \ R^{23}:=\sum 1\otimes X_a\otimes Y_a\in {\rm End}(V)^{\otimes 3}.$ } 
\begin{eqnarray*}\label{YBeq}
R^{12}R^{13}R^{23}=R^{23}R^{13}R^{12}\in {\rm End}(V)^{\otimes 3}.
\end{eqnarray*}
\end{thm}
The theorem suggests that the Yang-Baxter equation can be understood within the framework of the Stokes phenomenon.
In \cite{TLXu}\cite{Xu}, we show a similar result in the formal setting using a different approach. In the semiclassical setting, the theorem recovers Boalch's identification \cite{Boalch} of dual Poisson Lie groups with the moduli spaces of meromorphic connections (equipped with the Poisson structures from the irregular Atiyah-Bott construction \cite{Boalch2}), and its isomonodromy deformation recovers the Dubrovin connections of Frobenius manifolds \cite{Dubrovin} from enumerative geometry, as explained in \cite{Xu}.

\subsection{Universal $R$-matrix of $U_q({\rm sl}_n)$}
Note that the system \eqref{eq} can be decomposed into several rank $2$ systems. The Stokes matrices of a rank $2$ system has been computed (see \cite{BJL} Proposition 8). Thus one can compute explicitly the Stokes multiplier $R$ in a straightforward way. For example, if ${\rm dim}(V)=2$, the Stokes multiplier $R\in {\rm End}(V)^{\otimes 2}$ coincides with the evaluation of the universal $R$-matrix of $U_q({\rm sl}_2)$ (for $q=\frac{\pi i}{\kappa} $) in the natural representation $V$. We leave the computation of the Stokes multiplier in the general case along with other properties in a separate paper.

\subsection{Generality}
In this paper, for simplicity we only deal with the ${\rm dKZ}$ equation associated with ${\rm gl}_n$ and its natural representation. However, the Stokes phenomenon has been extended by Boalch \cite{Boalch1} from ${\rm GL}_n$ to any complex reductive Lie groups. Accordingly, our results can be generalized to a general ${\rm dKZ}$ equation \cite{FMTV} associated with a complex simple Lie algebra $\g$, an element $u$ in a Cartan subalgebra and a $\g$-module $V$. It is also interesting to study the monodromy/Stokes representation of other equations with irregular singularities appeared in mathematical physics and representation theory, for example the factorizable systems introduced by Cherednik \cite{Ch}. 

\subsection{Confluence of KZ equations} As pointed out by Etingof to us, the ${\rm dKZ}$ equation is a limit of the trigonometric ${\rm KZ}$ equation (see Section \ref{confluence}), which is the equation for conformal blocks of WZW model in genus $0$ (see e.g., \cite{Etingof} Section 3.8 for more details). Thus we expect that many theories related to the KZ equation should have a degeneration, which relates to irregular singularities and Stokes phenomenon. In particular, the ${\rm dKZ}$ equation may play a role in the theory of irregular conformal blocks, see e.g., \cite{GT}, and Theorem \ref{StokesYB}, relating Stokes matrices and Yang-Baxter relations, may have a gauge theoretic interpretation, in the spirit of \cite{CWY}. It also indicates that the ${\rm dKZ}$ equation may be related to the intertwining operators for affine Lie algebras. We will explore these possible relations in a future work.

\subsection*{Acknowledgements}
\noindent
I would like to thank Anton Alekseev, Philip Boalch, Pavel Etingof and Valerio Toledano Laredo for their useful discussions and comments on this paper. This work is partially supported by the Swiss National Science Foundation grants P2GEP2-165118 and P300P2-174284.

\section{Dynamical Knizhnik--Zamolodchikov equations}
In this section, we compute the monodromy of ${\rm dKZ}$ equations. In particular, the proofs of Theorem \ref{mainthm} and \ref{StokesYB} are given.
\subsection{{\rm dKZ} equations and braid groups}\label{secbraid}
Recall from the introduction that the ${\rm dKZ}_n$ equation for a function $F(z_1,...,z_n)$ of $n$ complex variables with values in $V^{\otimes n}$ is
\begin{eqnarray}\label{dKZ}
\kappa\frac{{\partial F }}{\partial z_i}= (u^{(i)}+ 
\sum_{j\ne i}\frac{ \Omega_{ij}}{z_i-z_j})\cdot F, \ \ \ \ \ \ i=1,...,n.
\end{eqnarray}
It is a local system over the configuration space
$X_n=\{(z_1,...,z_n)\in \mathbb{C}^n~|~z_i\ne z_j\}.$ Let the symmetric group $S_n$ act on $X_n$ by permutation of variables, and $X_n/S_n$ the quotient space. Then the fundamental group of $X_n/S_n$ is isomorphic to the braid group $B_n$ in $n$ strands.
Recall that $B_n$ has generators $b_1,...,b_{n-1}$ and relations
\begin{eqnarray}\label{braid}
b_ib_j &=&b_jb_i, \ \ \ |i-j| > 1,\\
b_ib_{i+1}b_i&=&b_{i+1}b_ib_{i+1}.
\end{eqnarray}
Actually choose a base point $z =(z_1,... ,z_n)$ such that $z_i\in\IR$, $z_1<z_2<\cdot\cdot\cdot<z_n$, then a homomorphism is given by 
$B_n\rightarrow \pi_1(X_n/S_n);~ b_i\mapsto the \ path \ in \ Figure \ 1.$
\begin{figure}[htb]
\[
  \begin{tikzpicture}
    [scale=0.5, baseline={([yshift=-.5ex]current bounding box.center)}]

    \filldraw (0,0) circle (2pt) node at (0,-0.5) {$z_1$};

    \filldraw (3,0) circle (2pt) node at (3,-0.5) {$z_2$};

    \filldraw (6,0) circle (2pt) node at (6,-0.5) {$z_{i}$};

    \filldraw (9,0) circle (2pt) node at (9.4,-0.5) {$z_{i+1}$};

    \filldraw (12,0) circle (2pt) node at (12,-0.5){$z_n$};

    \path[->, bend right=60] (6.2,-0.1) edge (8.8,-0.1);
    \path[->, bend right=60] (8.8,0.1) edge (6.2,0.1);
    \end{tikzpicture}
\]
   \caption{Transposition of $z_i$ and $z_{i+1}$ such that $z_{i+1}$ passes above $z_{i}$.}
\end{figure}
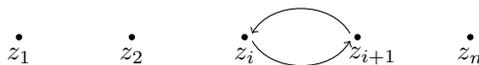

\subsection{Monodromy representation of ${\rm dKZ}$ equations}
Although the ${\rm dKZ}_n$ equation is in general not a local system over $X_n/S_n$, we can lift the loop $b_i$ in $X_n/S_n$ (an element of $B_n$) to a path in $X_n$, and consider the corresponding operator of holonomy along the path to obtain a representation of $\pi_1(X_n/S_n)\cong B_n$. To be more precise, we fix a base point $z\in D$, and denote by $M_i:V^{\otimes n}\rightarrow V^{\otimes n}$ the corresponding operator of holonomy along the path $b_i$ in Figure $1$. Let $T_i: V^{\otimes n}\rightarrow V^{\otimes n}$ be the permutation of the $i$-th and the $(i+1)$-th factors. In this way, we get
\begin{pro}
The map $b_i\mapsto  T_i\circ M_i(z)$ is a representation of the braid group $B_n$ in $V^{\otimes n}$, which does
not depend on the choice of $z$ (up to isomorphism).
\end{pro}
\pf It follows from the relation $b_ib_{i+1}b_i=b_{i+1}b_ib_{i+1}$ in the fundamental group of $X_n/S_n$. This definition depends on the choice of the base point. However, the monodromy operators $M_i(z')$ with respect to a new base point $z'$ can be obtained from
the old ones $M_i(z)$ by a conjugation.
\qed

\subsection{Solutions in asymptotic zones}\label{solsim}
Note that the solution of \eqref{dKZ} will have a leading power $e^{\frac{1}{\kappa}(\sum_{i=1}^n z_iu^{(i)})}$, which has different asymptotic behaviors as $z_i-z_j\rightarrow \infty$ in different patterns. This suggests the following construction, which can be seen as a special case of (a slight modification of) the isomonodromy deformation studied in Jimbo-Miwa-Ueno \cite{JMU} Section 3.

Let us consider the map $$P:\mathbb{C}^\times\times X_n\rightarrow X_n; \ P(z, \xi_1, . . . , \xi_n) = (z\xi_1, . . . , z\xi_n).$$
Then the pull-back of the ${\rm dKZ}$ equation \eqref{dKZ} under $P$ becomes 
\begin{eqnarray}\label{dKZ1}
&&\kappa\frac{{\partial F }}{\partial z}= (\sum_{i=1}^n \xi_iu^{(i)}+ 
\frac{ \sum_{i<j}\Omega_{ij}}{z})\cdot F, \\ \label{dKZ2}
&&\kappa\frac{{\partial F }}{\partial \xi_i}= (z u^{(i)}+ 
\sum_{j\ne i}\frac{ \Omega_{ij}}{\xi_i-\xi_j})\cdot F, \ \ \ for \ i=1,...,n.
\end{eqnarray}
For any $\xi$, the first equation has a formal fundamental solution taking the form\footnote{Here we use the fact the diagonal elements of $u$ are distinct, which implies the projection of $\sum_{i<j}\Omega_{ij}$ to the centralizer of $\sum_i \xi_iu^{(i)}$ is $\sum_{i<j}[\Omega]_{ij}$. See the Appendix.} (see e.g., \cite{Balser} Chapter 3) \begin{eqnarray}\label{formalsum}
\hat{F}=\hat{H} e^{\frac{z}{\kappa}(\sum_i\xi_iu^{(i)})}\prod_{i<j}z^{\frac{1}{\kappa}[\Omega]_{ij}}, \ \ \ {\it for} \ \hat{H}=1+H_1z^{-1}+H_2z^{-1}+\cdot\cdot\cdot.\end{eqnarray}
Here $H_i\in{\rm End}(V^{\otimes n})$ and $[\Omega]_{ij}$ denotes the action of $[\Omega]$ on the $i$-th and $j$-th factors of $V^{\otimes n}$. 

For any $k$, set $D_k= \{\xi\in \IR^n ~|~ \xi_1< \cdot\cdot\cdot <\xi_{k-1}< \xi_{k+1}<\xi_k <\cdot\cdot\cdot< \xi_n\}\subset X_n$. 
Since we assume the diagonal elements of $\frac{1}{\kappa}u$ are purely imaginary, for any fixed point $\xi\in D_k$ the right half plane $\IH_+$ is a Stokes sector of \eqref{dKZ1} (see Appendix). Thus there exists a function $H_{D_k}:\IH_+\rightarrow {\rm End}(V^{\otimes n})$ asymptotic to $\hat{H}$ as $z\mapsto\infty$ within $\IH_+$, and such that $F_{D_k}=H_{D_k} e^{\frac{z}{\kappa}(\sum_i \xi_iu^{(i)})}\prod_{i<j} z^{\frac{1}{\kappa}[\Omega]_{ij}}$ is a solution of \eqref{dKZ1}. Let us now consider the variation of $\xi$ in $D_k$.
\begin{pro}\label{solution}
Let $F_{k}(z;\xi):\IH_+\times D_k\rightarrow {\rm End}(V^{\otimes n})$ be the real analytic function given by
$$F_{k}(z;\xi):= F_{D_k}\cdot\prod_{i<j} (\xi_i-\xi_j)^{\frac{1}{\kappa}[\Omega]_{ij}}=H_{D_k} e^{\frac{z}{\kappa}(\sum_i \xi_iu^{(i)})}\prod_{i<j} (z\xi_i-z\xi_j)^{\frac{1}{\kappa}[\Omega]_{ij}}.$$ Then as a function on $\IH_+\times D_k$, $F_{k}(z;\xi)$ satisfies the systems \eqref{dKZ1} and \eqref{dKZ2}.
\end{pro}
\pf This statement follows from the isomonodromy deformation theory of equation \eqref{dKZ1} with respect to the variables $\xi_i's$, according to (a slight modification of) \cite{JMU} Section 3. In the following we include a complete proof for the readers' convenience.

Set $\nabla_z=\frac{d}{dz}-(\sum_{i=1}^n \xi_iu^{(i)}+ 
\frac{ \sum_{i<j}\Omega_{ij}}{z})$ and $\Phi=\frac{1}{\kappa}\sum_i(z u^{(i)}+ \sum_{j\ne i}\frac{ \Omega_{ij}}{\xi_i-\xi_j})d\xi_i$. From the compatibility of the equations \eqref{dKZ1} and \eqref{dKZ2}, we have $\nabla_z(d_\xi F_k-\Phi F_k)=0$, where $d_\xi$ denotes the exterior differentiation with respect to parameters $\xi_i's$. It implies that there exists a matrix $B_i$ of 1-forms independent of $z$ such
that $d_\xi F_k-\Phi F_k=F_k B_k$. To show $B_k=0$, we compare the expansion of the both sides of this equation at $z=\infty$. 

Firstly, the formal sum $\hat{H}=1+H_1z^{-1}+H_2z^{-2}+\cdot\cdot\cdot$ in \eqref{formalsum} satisfies $$\frac{dH}{dz}+H\cdot \frac{1}{\kappa}(\sum_{i=1}^n \xi_iu^{(i)}+ 
\frac{ \sum_{i<j}[\Omega]_{ij}}{z})=\frac{1}{\kappa}(\sum_{i=1}^n \xi_iu^{(i)}+ 
\frac{ \sum_{i<j}\Omega_{ij}}{z})\cdot H.$$
Comparing the coefficients of $z^{-1}$, we find out $H_1(\xi)$ satisfies \begin{eqnarray}
\label{simH}
[\sum_{i=1}^n \xi_iu^{(i)}, H_1]=\sum_{i<j}\Omega_{ij}-\sum_{i<j}[\Omega]_{ij}.\end{eqnarray}
Since $d_\xi F_k\cdot F_k^{-1}=d_\xi H_{D_k}\cdot H_{D_k}^{-1}+H_{D_k}\left(\sum_i \frac{z}{\kappa}u^{(i)}d\xi_i +\sum_{i<j}\frac{[\Omega_{ij}]d(\xi_i-\xi_j)}{\kappa(\xi_i-\xi_j)}\right)H_{D_k}^{-1}$ and $H_{D_k}\sim \hat{H}=1+H_1z^{-1}+\cdot\cdot\cdot$, we get

\begin{eqnarray}\label{sim}d_\xi F_k\cdot F_k^{-1}-\Phi\sim O(z^{-1}) \  \ at \  \ z=\infty \ \ \ in \ \ \IH_+.\end{eqnarray}
Secondly, since $F_k B_k F_k^{-1}\sim \prod_{i<j} (z\xi_i-z\xi_j)^{\frac{[\Omega]_{ij}}{\kappa}} e^{\frac{z}{\kappa}(\sum_{i} \xi_iu^{(i)})} B_i e^{-\frac{z}{\kappa}(\sum_i \xi_iu^{(i)})}\prod_{i<j} (z\xi_i-z\xi_j)^{-\frac{[\Omega]_{ij}}{\kappa}} $ in the supersector $\widehat{\IH}_+=\{\rho e^{i\theta}~|~\rho>0, -\pi<\theta<\pi\}$ (see Appendix) and the exponentials dominate, $B_i$ must be a diagonal matrix. (Otherwise the non-vanishing off-diagonal element of  $e^{\frac{z}{\kappa}(\sum_{i} \xi_iu^{(i)})} B_i e^{-\frac{z}{\kappa}(\sum_i \xi_iu^{(i)})}$ would grow exponentially, for the central angle of $\widehat{\IH}_+$ is
larger than $\pi$.) Thus 
$$ F_k B_k  F_k^{-1}\sim B_i +O(z^{-1})  \  \ at \  \ z=\infty \ \ \ in \ \ \IH_+.$$
Comparing it with \eqref{sim}, we obtain that $B_i=0$. It proves that $F_k(z;\xi)$ is a solution of the linear systems \eqref{dKZ1} and \eqref{dKZ2}. \qed

\vspace{3mm}
We assume initially that $z_1,...,z_n$ are real in the ${\rm dKZ}_n$ equation \eqref{dKZ}. Proposition \ref{solution} enables us to construct unique (therefore canonical) solutions with prescribed asymptotics at infinity associated to the domains $D_0=\{z\in \IR^n~|~z_1<z_2<\cdot\cdot\cdot <z_n\}$ and $D_i=\{z~|~z_1<\cdot\cdot\cdot<z_{i-1}<z_{i+1}<z_i<\cdot\cdot\cdot z_n\}$ for $i=1,..., n-1.$

\begin{defi}
We denote by $F_i$ (resp. $F_0$) the canonical solution of \eqref{dKZ} associated to the domain $D_i$ (resp. $D_0$).
\end{defi}

\subsection{Braid groups representations}\label{Monorep}
To compute the monodromy of equation \eqref{dKZ}, we take the infinite base point $z=(z_1,...,z_n)$ in $D_0$ (i.e., $z_i\ll z_{i+1}$ for any $i$) \footnote{See \cite{Etingof} Section 8.4 for the discussion for the monodromy with respect to an infinite base point in the case of ${\rm KZ}$ equations. In a similar way, one checks this method is valid in our situation.}. Then the induced braid group representation is
$$\pi_1(X_n/S_n)\rightarrow {\rm End}(V^{\otimes n}) \ ; \ b_i\mapsto F_i \cdot F_0^{-1},$$
where $b_i's$ are generators of $B_n$, and the image is the ratio of the canonical solutions $F_0$ and $F_i$ taken in $D_i$ (after $F_0$ has been analytic continued to $D_i$ along the path $b_i$ in Figure 1.). 

\begin{thm}\label{monodromrep}
The monodromy representation of the ${\rm dKZ}_n$ equation is given by
$$\pi_1(X_n/S_n)\cong B_n\rightarrow {\rm End}(V^{\otimes n}); \ b_i\mapsto T_i\circ R^{i,i+1},$$
where $R^{i,i+1}$ is the action of the Stokes multiplier $R$ on the $i$-th and $(i+1)$-th factors of $V^{\otimes n}$.
\end{thm}
\pf For $n=2$ case. Recall that the ${\rm dKZ}_2$ equation is equivalent to the equation \eqref{eq}
\begin{eqnarray*}
\kappa\frac{dF}{dz}= (u^{(1)}+ \frac{\Omega_{12}}{z})\cdot F.
\end{eqnarray*}
Thus the canonical solutions associated with the asymptotic zones $z_1-z_2\gg0$ and $z_1-z_2\ll0$ are $$F_0(z_1,z_2)=e^{\frac{z_2}{\kappa}(u^{(1)}+u^{(2)})}Y_+(z_1-z_2), \ \ \ F_1(z_1,z_2)=e^{\frac{z_2}{\kappa}(u^{(1)}+u^{(2)})}Y_-(z_1-z_2),$$
where $Y_+(z)$ and $Y_-(z)$ are solutions of the equation \eqref{eq} given in the Introduction. Furthermore, analytic continuation of $F_0$ along the path $b_1$ amounts to the continuation of $Y_+(z)$ to the left half plane in a counterclockwise direction. Thus being aware of the half formal monodromy, the monodromy representation is $$\pi_1(X_2/S_2)\cong B_2\rightarrow {\rm End}(V^{\otimes 2})~;~b_1\mapsto T_1\circ R.$$
Here $R$ is the Stokes multiplier given in \eqref{Stokesfactor}.

For general $n$ case. We need to compute the ratio of the solutions $F_0$ and $F_i$. To this end, considering the asymptotic zone $z_{j+1}-z_j\gg 0$ for any $j$, and fixing the variables $z_1,...,z_{i-1},z_{i+2},...,z_n$ of the ${\rm dKZ}$ equation, we are left with the equation
\begin{eqnarray*}
&&\kappa\frac{\partial F}{\partial z_i}= (u^{(i)}+ \sum_{j\ne i}\frac{\Omega_{ij}}{z_i-z_j})\cdot F,\\
&&\kappa\frac{\partial F}{\partial z_{i+1}}= (u^{(i+1)}+ \sum_{j\ne i+1}\frac{\Omega_{i+1,j}}{z_{i+1}-z_j})\cdot F.
\end{eqnarray*}
If we further restrict to the asymptotic zone $|\frac{z_i-z_{i+1}}{z_i-z_{j}}|, |\frac{z_i-z_{i+1}}{z_{i+1}-z_{j}}|\ll 1$ for any $j\ne i, i+1$, the above equation is approximated by 
\begin{eqnarray*}
&&\kappa\frac{\partial F}{\partial z_i}= (u^{(i)}+ \frac{\Omega_{i,i+1}}{z_i-z_{i+1}})\cdot F,\\
&&\kappa\frac{\partial F}{\partial z_{i+1}}= (u^{(i+1)}+ \frac{\Omega_{i,i+1}}{z_{i+1}-z_i})\cdot F.
\end{eqnarray*}
In view of their asymptotics, it reduces the computation of the monodromy to the $n=2$ case. In particular, the ratio of the solutions $F_0$ and $F_i$ is given by 
$R^{i,i+1}.$
\qed

\vspace{3mm}
As a corollary, we have
\begin{thm}
The Stokes multiplier $R=e^{\frac{\pi i}{\kappa} [\Omega]} S_+$ satisfies the Yang-Baxter equation.
\end{thm}
\pf It follows from Theorem \ref{monodromrep} and the braid relation \eqref{braid}.
\qed

\vspace{3mm}
In the computation given in \eqref{Monorep}, if we replace the infinite base point in $D_0$ by the infinite base point $z=(z_1,...,z_n)$ such that $z_i\in\IR$, $z_i\gg z_{i+1}$, then the corresponding braid group representation will be given by
$$\pi_1(X_n/S_n)\cong B_n\rightarrow {\rm End}(V^{\otimes n}); \ b_i\mapsto T_i\circ R_-^{i,i+1},$$
where $R_-:=e^{\frac{\pi i}{\kappa}[\Omega]} S_-$ is another Stokes multiplier of equation \eqref{eq} and $R_-^{i,i+1}$ stands for the action on the $i$-th and $(i+1)$-th factors of $V^{\otimes n}$. In particular, the Stokes multiplier $R_-$ also satisfies the Yang-Baxter equation.

\section{Discussions}
\subsection{Factorization property}\label{advantages}
The main advantage of the monodromy representation of ${\rm dKZ}$ equations, comparing to the similar results about $\rm KZ$ equations (see e.g., \cite{Etingof} Section 8), is the stronger factorization property (Theorem \ref{monodromrep}). Recall from \cite{Etingof} Section 8.5 that the monodromy representation of ${\rm KZ}_n$ equation is determined by the $n=3$ case (which reduces to)
\begin{eqnarray*}
\kappa\frac{dF}{dz}=(\frac{\Omega_{12}}{z}+\frac{\Omega_{23}}{z-1})\cdot F.
\end{eqnarray*}
Its monodromy involves two data: the monodromy matrix at a simple pole and the connection matrix between two poles (corresponding respectively to the R-matrix and the associator in Drinfeld's quasi-Hopf algebras \cite{Drinfeld}). 

While following Theorem \ref{monodromrep}, the monodromy representation of ${\rm dKZ}_n$ equation is determined by the $n=2$ case. Its monodromy only involves one data: the Stokes matrix. It is due to the fact that the irregular singularities dominate in ${\rm dKZ}$ equation, thus the connection matrices (associator) between regular singularities don't show up any more. The similar idea appears in the construction of Drinfeld twists by Alekseev-Torossian \cite{AT} and Toledano Laredo \cite{TL}.

It seems that as a payback of the stronger factorization property, one has to work with the Stokes phenomenon or the Stokes matrices, which in general are highly transcendental. However, the system \eqref{eq} can be decomposed into several rank $2$ systems, in which case the Stokes matrices have been computed (see \cite{BJL} Proposition 8). In particular, if ${\rm dim}(V)=2$, the Stokes multiplier $R\in {\rm End}(V)^{\otimes 2}$ coincides with the evaluation of the universal $R$-matrix of $U_q({\rm sl}_2)$ (for $q=\frac{\pi i}{\kappa} $) in the natural representation.

\subsection{Stokes representation, isomonodromy and Stokes factors}
The monodromy representation obtained in Theorem \ref{monodromrep} is closely related to Boalch's Stokes representation \cite{Boalch2} and the isomonodromy deformation \cite{Boalch1, JMU} of the equation \eqref{dKZ1}. Furthermore, the decomposition of the Stokes multiplier $R$ into Stokes factors (see \cite{BJL} Section 4) is translated to the multiplicative property of the universal $R$-matrix of $U_q({\rm sl}_n)$ \cite{FR}. We will leave the computation of $R$ along with these properties in a forthcoming paper. 

\subsection{Confluence of ${\rm KZ}$ equations}\label{confluence}
First note that the results in the paper can be generalized to the {\rm dKZ} equations associated to any complex simple Lie algebra. 
Let $\g$ be such a Lie algebra with a Cartan subalgebra $\frak h$ and a root system $\Delta=\Delta_+\sqcup \Delta_-$. For $\alpha\in\Delta$, we choose generators $e_\alpha$ of the root subspaces $\g_\alpha$ such that $(e_\alpha,e_{-\alpha})=1$, and choose $\{x_i\}$ an orthogonal basis of $\h$. We set $$[\Omega]=\sum x_i\otimes x_i, \ \ \ \ \Omega^+=\frac{1}{2}[\Omega]+\sum_{\alpha\in \Delta_+}e_\alpha \otimes e_{-\alpha}, \ \ \ \Omega^-=\frac{1}{2}[\Omega]+\sum_{\alpha\in \Delta_+}e_{-\alpha} \otimes e_\alpha.$$ Define the Casimir element and the trigonometric $r$-matrix (see e.g., \cite{Etingof} Section 3.8 or \cite{EV}) by $$\Omega=\Omega^++\Omega^-, \ \ \ \ \ \ r(z)=\frac{\Omega^+z+\Omega^-}{z-1}.$$
Then the trigonometric ${\rm KZ}_n$ equation associated to a $\g$-module $V$ and a regular element $u\in \frak h_{\rm reg}$ is 
\begin{eqnarray*}\label{TKZ}
\kappa z_i \frac{\partial F}{\partial z_i}=(t u^{(i)}+\sum_{j\ne i} r(z_i/z_j)^{ij})\cdot F, \ \ \ {\it for} \ i=1,...,n,
\end{eqnarray*}
where $F(z_1,...,z_n)$ is valued in $V^{\otimes n}$, $r^{ij}$ and $u^{(i)}$ denote $r$ acting in the $i$-th and $j$-th factors of the tensor product and $u$ acting in the $i$-th factor, and $t$ is a complex parameter.

In terms of $s_i=t (1+z_i)$, the above ${\rm KZ}_n$ equation becomes
\begin{eqnarray*}
\kappa \frac{\partial F}{\partial s_i}=\left(\frac{t u^{(i)}}{s_i-t}+\sum_{j\ne i} (\frac{\Omega_{ij}}{s_i-s_j}-\frac{\Omega_{ij}^-}{s_i-t})\right)\cdot F, \ \ \ {\it for} \ i=1,...,n.
\end{eqnarray*}
When $t\rightarrow \infty$, it approaches to the ${\rm dKZ}_n$ equation
\begin{eqnarray*}
\kappa \frac{\partial F}{\partial s_i}=(-u^{(i)}+\sum_{j\ne i}\frac{\Omega_{ij}}{s_i-s_j})\cdot F, \ \ \ {\it for} \ i=1,...,n.
\end{eqnarray*}
Therefore, at the level of equations, the ${\rm dKZ}_n$ \eqref{dKZ} is a limit of the ${\rm KZ}_n$ \eqref{TKZ}. At the level of solutions, the limit is related to the fact that the Kummer's function (or confluent hypergeometric function) is a limit of the hypergeometric function. Thus we expect that many theories related to the ${\rm KZ}$ equations should have a degeneration, which is coupled with irregular singularities and Stokes phenomenon.

\section{Appendix}
\subsection{Canonical solutions and Stokes matrices}
Let us consider the meromorphic linear system
\begin{eqnarray}\label{nabla}
\frac{dF}{dz}=(\lambda+\frac{A}{z})\cdot F,
\end{eqnarray}
where $F(z)\in \mathbb{C}^m$, $\lambda={\rm diag}(\lambda_1,...,\lambda_m)$ is a diagonal matrix, and $A\in {\rm gl}_m$ an arbitrary matrix. Thus $z=\infty$ is an irregular singularity, and the equation in general has only a formal solution around $\infty$ taking the form 
\begin{eqnarray}\label{fomralsol}
\hat{F}(z)=\hat{H}(z)z^{[A]} e^{z \lambda}, \ \ \ {\it for} \ \ \hat{H}(z)=1+H_1z^{-1}+H_2z^{-2}+\cdot\cdot\cdot.
\end{eqnarray}
Here $[A]$ takes the projection of $A$ to the centralizer of $\lambda$ in ${\rm gl}_m$. In particular, if $\lambda$ has distinct diagonal elements, $[A]$ takes the diagonal part of $A$. 

Such a formal power series is, in fact, asymptotic to canonical holomorphic solutions $F_i(z)$ (given by resummation methods, see e.g., \cite{Balser} Chapter 4-6) in certain different sectors ${\rm Sect}_i$ in the complex plane. The discontinuity of asymptotic expansions is known as the Stokes phenomenon, and is measured by the so called Stokes matrices which connect the different solutions $F_i$ with the fixed asymptotic expansion $\hat{F}$ in the various sectors. We give more details in the following.

\begin{defi}\label{Stokesrays}
The {\it anti-Stokes rays} of the equation \eqref{nabla} are the directions
along which $e^{(\lambda_i-\lambda_j)z}$ decays most rapidly as $z\mapsto \infty$ for some $i\ne j$. The {\it Stokes sectors}
are the open regions of $\mathbb{C}$ bounded by two adjacent anti-Stokes rays.
\end{defi}
On each Stokes sector ${\rm Sect}_i$ bounded by two Stokes rays $d_i$ and $d_{i+1}$, there is a canonical
solution $F_i$ of \eqref{nabla} with prescribed asymptotics on the supersector $\widehat{{\rm Sect}_i}=(d_i-\frac{\pi}{2},d_{i+1}+\frac{\pi}{2})$.
In particular, the following result can be found in e.g., \cite{Balser} Chapter 8 or \cite{BJL}\cite{MR}.
\begin{thm}\label{jurk}
On ${\rm Sect}_i$, there
is a unique (therefore canonical) holomorphic function $H_i:{\rm Sect}_i\to {\rm GL}_n$ such that the function
\[F_i=H_i e^{z\lambda} z^{[A]}\]
satisfies equation \eqref{nabla}, and $H_i$ can be analytically continued to $\widehat{{\rm Sect}_i}$ and then $H_i$ is asymptotic to $\hat{H}$ at $z\mapsto \infty$ within $\widehat{{\rm Sect}_i}$. 
\end{thm}

Suppose we are given a Stokes sector ${\rm Sect}_0$ (with a chosen branch of ${\rm log}(z)$ on it) and the opposite sector ${\rm Sect}_l$. 
\begin{defi}
The {\it Stokes matrices} of the equation \eqref{nabla} (with respect to
to ${\rm Sect}_0$) are the matrices $S_\pm$ determined by
\[F_l=F_{0}\cdot S_+, \ \ \ \ \ 
F_{0}=F_l\cdot e^{2\pi i [A]} S_-
\]
where the first (resp. second) identity is understood to hold in ${\rm Sect}_l$
(resp. ${\rm Sect}_0$) after $ F_0$ (resp. $ F_{l}$)
has been analytically continued counterclockwise. 
\end{defi}
We remark that the Stokes matrices $S_\pm$ will in general depend on the irregular data $\lambda$ in \eqref{nabla}. Such dependence was studied by many authors, see e.g., \cite{Boalch1}\cite{JMU}.

\Addresses

\end{document}